\documentclass[a4paper,12pt]{article}
\usepackage[cp1251]{inputenc}
\usepackage{graphics}
\usepackage{amssymb,amsmath}
\usepackage[english]{babel}
\usepackage{indentfirst}
\begin{document}

\bigskip
\centerline {\bf SOLAR ACTIVITY INDICES IN 21,22 AND 23 CYCLES }

\bigskip

\centerline {E.A.Bruevich $^{a}$, G.V. Yakunina $^{b}$}

\bigskip

\centerline {Sternberg Astronomical Institute, Moscow, Russia}\

\centerline
{E-mail:$^a${red-field@yandex.ru},$^b${yakunina@sai.msu.ru}}\

\bigskip

The monthly average values of the main solar activity indices such
as Wolf numbers (W), 10,7 cm radio flux $F_{10,7}$, 0,1-0,8 nm
background, total solar irradiance and Mg II UV-index (280 nm core
to wing ratio) were studied for 23 activity cycle. The correlation
coefficients of linear regression for main solar activity indices
versus $F_{10,7}$ were analyzed for every year of 23 activity cycle.
All the correlation coefficients for 23 activity cycle have the
minimum values in 2001. We also analyzed the connection between
monthly average values of the most popular indices W and $F_{10,7}$
for 21, 22 and 23 solar activity cycles. The values of every year
linear regression correlation coefficients for W versus $F_{10,7}$
show the cyclic variations with period closed to half path of
so-called 11-year period of solar activity.


\vskip12pt {\it\bf Introduction}
\vskip15pt

We analyze the main
solar activity indices variations at 11-year time scale. These
indices are: Wolf numbers W, 10,7 cm radio flux $F_{10,7}$, the
X-ray 0,1-0,8 nm background, the UV Mg II 280 nm core to wing ratio
 and total solar irradiance.
Most of the observational data we used in our issue were published
in Solar-Geophysical Data bulletin. Solar-Geophysical Data was a
monthly bulletin issued from 1955 to 2009 in one or two parts
(published by NOAA National Geophysical Data Center) that provides
the scientific community with a variety of solar-terrestrial data.

The Wolf numbers is a very popular solar activity index: the
  series of Wolf numbers observations continue more than two hundred
years.

The solar radio microwave flux at wavelengths 10,7 cm $F_{10,7}$ has
also the longest running series of observations (it's nominally an
absolute flux, measured in units $10^{-22}$ Watt per square meter
per Hertz) started in 1947 in Ottawa, Canada and maintained to this
day at Pentiction site in British Columbia.

This emission comes from high part of the chromosphere and low part
of the corona. It has two different sources: thermal bremsstrahlung
(due to electrons radiating when changing direction by being
deflected by other charged participles) and gyro-radiation (due to
electrons radiating when changing direction by gyrating around
magnetic fields lines) [1]. These mechanisms give rise to enhanced
radiation when the temperature, density and magnetic fields are
enhanced. So $F_{10,7}$ is a good measure of general solar activity.

The Mg II 280 nm is important solar activity indicator of radiation.
It has been shown that the Mg II index derived from daily solar
observations of the core-to-wing ratio of the Mg II doublet at 279,9
nm provides a good measure of the solar UV variability and can be
used as a reliable proxy to model extreme UV variability during the
solar cycle [2,3,4].

The Mg II observation data were obtained from several satellite
(NOAA, ENVISAT) instruments. NOAA started in 1978 (during 21,22 and
first part of 23 solar activity cycles), ENVISAT was launched on
2002 (last part of 23 solar activity cycle). Comparison of the NOAA
and ENVISAT  Mg II index observation data shows that both the Mg II
indexes agree to within about 0,5\%. The correlation coefficient is
about 0,995 - very high. We used the ENVISAT  Mg II index
observation data for the last part of 23 solar activity cycle
published in Solar Geophysical Data bulletin and NOAA Mg II index
observation data for the first part of 23 solar activity cycle
published in [2].

Data of GOES observations of the X-ray 0,1-0,8 nm background for 23
cycle were taken from Solar Geophysical Data bulletin. This
permanent monitoring of solar disk in 0,1-0,8 nm is sufficient
enough not only for flares observations and  flares prediction but
0,1-0,8 nm background observational data is a good activity index of
solar corona activity without flares.

Data of total solar irradiance SOHO observations for 23 cycle were
taken from Solar Geophysical Data bulletin. Earlier from 1985 to
2000 total solar irradiance  was observed by Earth Radiation Budget
Satellite (EBRS). The values of total solar flux vary from 1364 to
1367 Watt per square meter for 21,22 and 23 cycles.

\begin{figure}[h!]
 \centerline{\includegraphics{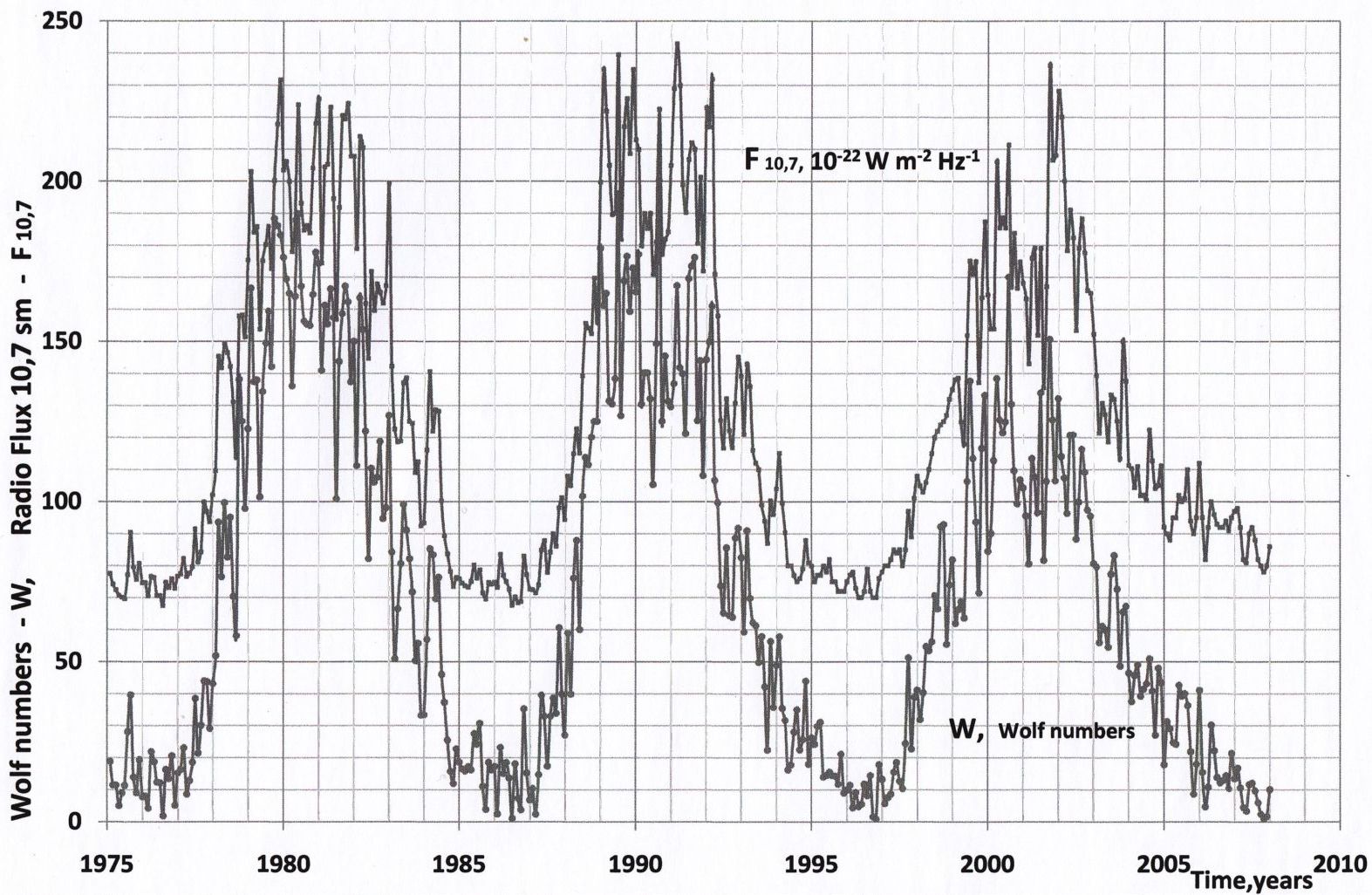}}
 \caption{The Wolf numbers and 10,7 cm flux $F_{10,7}$}(monthly average values)
 in 21, 22 and 23 solar activity cycles\label{Fi:Fig1}
\end{figure}

 The 11-year fluxes variation are widely spread phenomenon for F, G and K stars
(not only for Sun). The activity indices variations for solar-type
stars were studied at Mount Wilson observational program during 45
years, from 1965 to present time [8].

These observations of the sensitive chromospheric activity
indicators (radiative fluxes at the centers of the H and K emission
lines of Ca II - $396,8$ and $393,4$ nm respectively) show that
about 30 \% of Mount Wilson project stars are characterized by the
regular cyclic activity on the 11-year time scale [9].

Figure 1 demonstrates monthly average values of Wolf numbers and
$F_{10,7}$ in 21, 22 and 23 solar activity cycles. The Wolf numbers
and $F_{10,7}$ fluxes show the pronounced determined two maximum in
all the activity cycles.

The most expressive two maximum form (at the first half of 2000 and
at the end of 2001 years) we see in 23 solar cycle. Also we see here
the local minimum in the middle of the 2001. The 23 solar activity
cycle was the medium-sized cycle with respect to values of maximum
amplitudes of main solar monthly average activity indexes variations
[5,6].

It is probable that the two-maximum shape of 11-year cycling solar
indices variations cased by quasi-biennial modulation of the 11-year
time scale solar activity cycle. The quasi-biennial fluxes variation
are widely spread phenomenon for star's (not only for Sun)
observations too. The periods of star's fluxes variations on the
quasi-biennial time scale vary from 2,2 to 3,5 years [7].

It has been shown that there exist the close connection between
photospheric and coronal fluxes variations for solar-type stars of
F, G, K and M spectral classes with active atmospheres [10].

\vskip15pt
\vskip15pt
\vskip15pt {\it\bf 1. Activity indices in 23
cycle}
\vskip15pt

\begin{figure}[h!]
 \centerline{\includegraphics{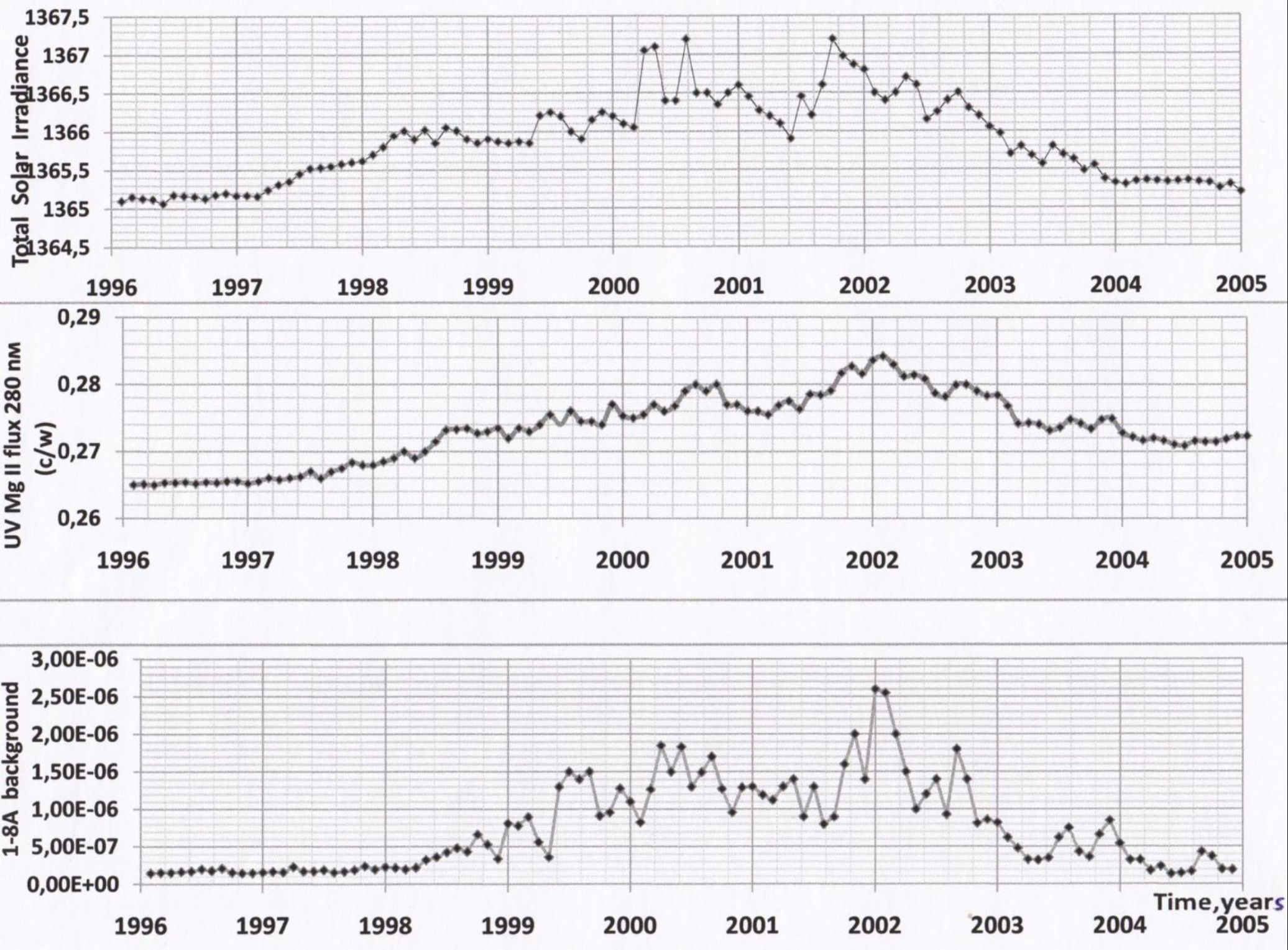}}
 \caption{The total solar irradiance (Watt per square meter), UV MgII 280 nm flux (c/w ratio) and 1-8A background
 (Watt per square meter)in 23 solar activity cycle (monthly average values) according to
 observations data published in Solar-Geophysical Data bulletin and [2]}\label{Fi:Fig2}
\end{figure}

At Figure 2 we demonstrate monthly average values of the total solar
irradiance, UV MgII 280 nm flux (c/w ratio) and 1-8A background in
23 solar activity cycle. We've made a comparison of  maximum values
occurrence time of these three indexes. We also see existence of two
maximum shape of all activity indices in 23 cycle.

\vskip15pt
\vskip15pt
\begin{center}
\centerline{Table 1}
 \begin{tabular}{|c|c|c|c|c|} \hline
  activity indices &1  max time  &2 max time&variation amplitude&flux formation\\ \hline
  Wolf numbers     & Apr-May 2000& Nov 2001  &  from 1 to 175      &photosphere  \\ \hline
  $F_{10,7}$       & Apr-May 2000& Dec 2001  &  from 70 to 240     &lower Corona \\ \hline
  MgII (280 nm c/w)& Jul-Sep 2000& Jan 2002  &from 0,265 to 0,284  &chromosphere  \\ \hline
  Total Solar flux &Mar-July 2000& Sep 2000  & from 1365 to 1367,3 &all atmosphere\\ \hline
  0,1-0,8 nm b/gr  & Mar-Apr 2000& Jan 2002  &$10^{-9}$ - $2,6\cdot10^{-6}$&corona    \\ \hline

\end{tabular}
\end{center}

\vskip15pt

 Table 1 demonstrates 1-st
and 2-nd time of maximum occurrence of W, $F_{10,7}$, the total
solar irradiance, UV MgII 280 nm flux (core to wing ratio) and 1-8A
background in 23 solar activity cycle. A comparison of maximum
values occurrence time of these five indexes shows that there are
displacements in both maximum occurrence time.

Table 1 also demonstrates that these five indexes describe changes
in physical conditions at all the levels of solar atmosphere for the
different phases of 11-year activity cycle. It's interesting to note
that amplitudes of fluxes variations in activity cycle differ much
for these indices. For total solar irradiance (generally
characterized by the photosphere flux contribution) the maximum
amplitude variations are some parts of percent. At the other hand
the 0,1-0,8 nm background  flux (characterized by the solar corona
contribution) changed from cycle's minimum to maximum more than a
thousand times. Note than the Wolf numbers cycling variations don't
have the direct physical interpretation because of W values
calculate only numbers of spots and spot groups. So the Wolf numbers
are not the real solar fluxes from photosphere but  anyway it's the
very important index of solar activity because of existence of long
homogeneous series of Wolf numbers observations.

\vskip15pt {\it\bf 2. The connection between activity indices in 23
cycle}
\vskip12pt

 The most popular solar activity index $F_{10,7}$
correlates well with Wolf numbers (see Fig 1). We estimated here the
linear connection between $F_{10,7}$ and other three indices: Mg II
(280 nm), total solar irradiance and 0,1-0,8 nm background.

The study of the indices variations connection among themselves and
particularly with $F_{10,7}$ indicates than there is a close linear
relationships between these activity indices and $F_{10,7}$. These
relationships were confirmed by the high-correlation. Usually the
correlation coefficients $K_{corr}$ is more than 0,7. According to
our calculations the highest values of  correlation coefficient
$K_{corr}$ we see between W and $F_{10,7}$  but correlation
coefficient $K_{corr}$
 between 0,1-0,8 nm background flux and $F_{10,7}$ is minimal of
 all correlations coefficients determined here.

\vskip12pt \centerline{Table 2}
\vskip12pt \centerline{The
correlation coefficients $K_{corr}$ in 23 solar activity cycle}
\vskip12pt
\begin{center}
 \begin{tabular}{|c|c|c|c|c|} \hline
  activity indices        & growth phase          & decrease phase & cycle maximum  & all the cycle \\ \hline
  W - $F_{10,7}$          &0.91951471             &0.96151775   &0.74290243      &0.93983805     \\ \hline
  Mg II- $F_{10,7}$       &0.96305749             &0.96482504   &0.75747438      &0.87903753     \\ \hline
  Tot Sol Irr - $F_{10,7}$&0.87991344             &0.94938041   &0.7435922       &0.92010067     \\ \hline
  0,1-0,8nm - $F_{10,7}$  & 0.8992561             &0.81403471   &0.7732229       &0.81241226     \\ \hline

\end{tabular}
\end{center}
\vskip12pt

 When studied five activity indices in 23 solar activity
cycle we separate out growth  cycle phase (from Oct 1997 to Nov
1997), cycle maximum phase (from Nov 1997 to Jul 2002) and decrease
cycle phase (from
 Jul 2002 to Jan 2006). Two maximum values of activity indices we can see at Figure 1 and Table 1.
In [5] it has been noted that $F_{10,7}$ and total solar irradiance
have the lowest values from 2007 to 2009 all over observation time.

We've calculated yearly values of $K_{corr}$ of linear dependence
between 4 solar activity indices and $F_{10,7}$ for 23 solar
activity cycles. Figure 3 demonstrates the yearly values of
$K_{corr}$ variations during 23 solar cycle. We see that all the
 $K_{corr}$ values of linear dependence between 4 activity
indices and $F_{10,7}$ have the maximum values at growth and at
decrease cycle phases. Also we see that in 2001 for all the activity
indices yearly values of $K_{corr}$ has a minimum value.

\begin{figure}[h!]
 \centerline{\includegraphics{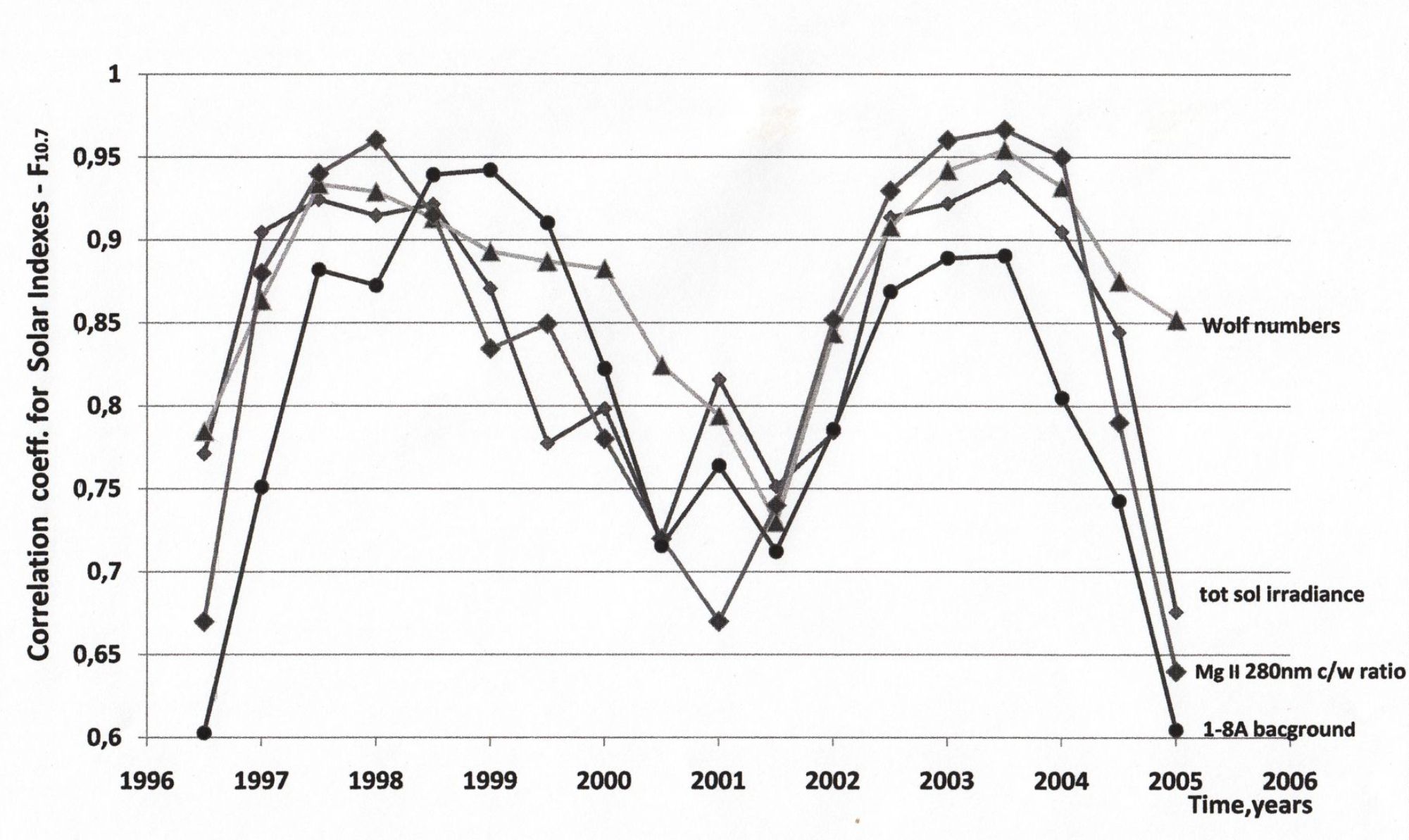}}
 \caption{Yearly correlation coefficients of linear regression $K_{corr}$ between the Wolf numbers,
  the total solar irradiance, UV Mg II 280 nm flux (c/w ratio), 1-8A background
  and $F_{10,7}$ in 23 solar activity cycle}\label{Fi:Fig3}
\end{figure}

\vskip15pt {\it\bf 3. The connection between W and $F_{10,7}$ in
21,22,23 solar activity cycles}
\vskip12pt

As we note earlier Wolf numbers are very useful because of existence
of long homogeneous observations series. The  $F_{10,7}$ series of
observations (data of Pentiction observatory in British Columbia)
are homogeneous for the 21,21 and 23 solar activity cycles too. So
we can study the linear dependence between W and $F_{10,7}$ in
21,22,23 solar activity cycles and calculated yearly values of this
linear regression coefficients $K_{corr}$.

At Figure 4 we demonstrate yearly values of correlation coefficients
of linear regression between W and $F_{10,7}$ in 21,22 and 23 solar
activity cycles. We see the cyclic behavior yearly values of
$K_{corr}$ of linear regression between W and $F_{10,7}$ during
21,22 and 23 solar activity cycles. We can estimate the value of
period of $K_{corr}$ cyclic variations as 5,5 years approximately.

\begin{figure}[h!]
 \centerline{\includegraphics{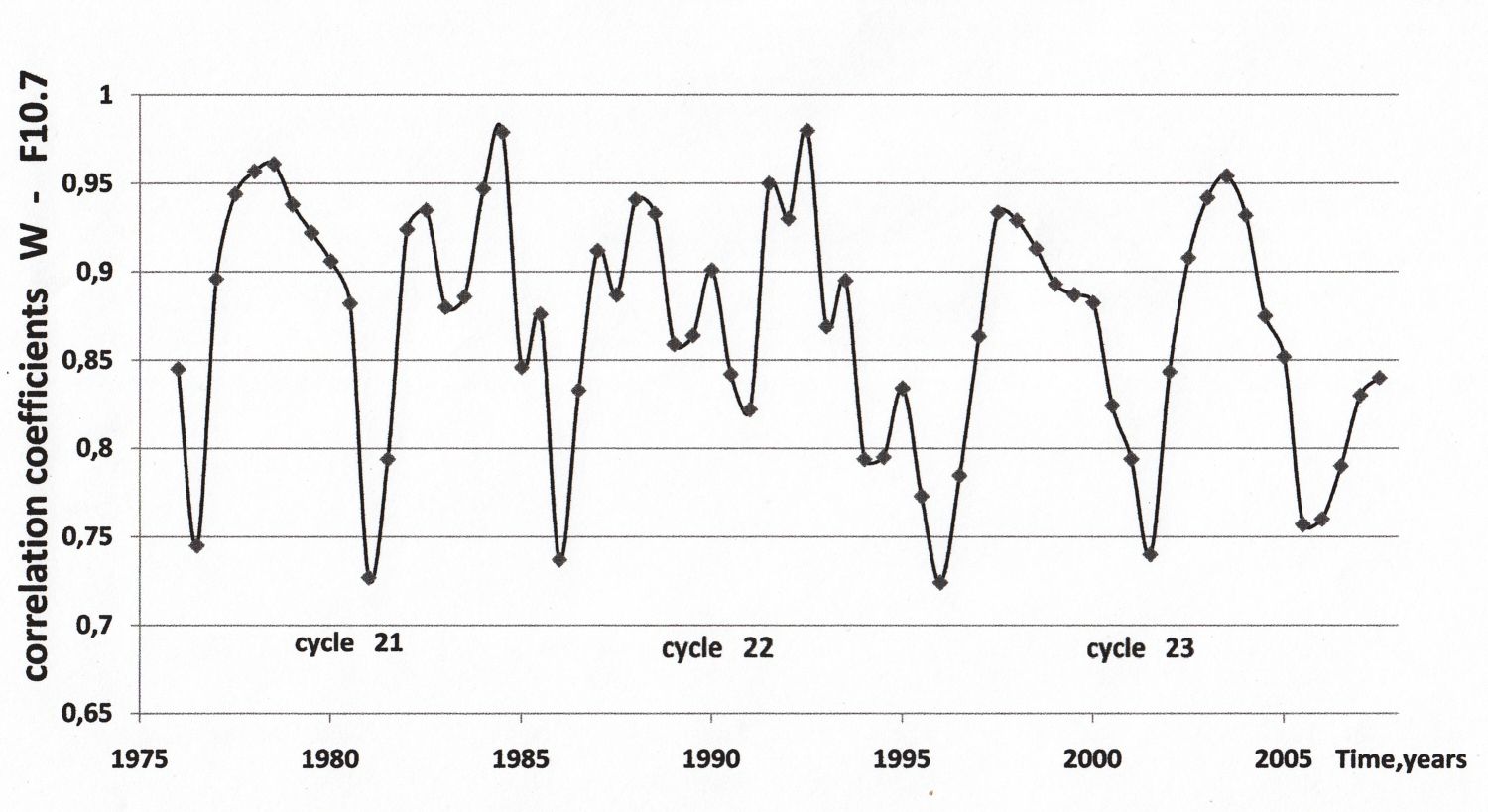}}
 \caption{Yearly correlation coefficients of linear regression $K_{corr}$ between Wolf numbers and $F_{10,7}$
 in 21,22 and 23 solar activity cycles}\label{Fi:Fig4}
\end{figure}

\vskip15pt {\it\bf Conclusions}
\vskip12pt

Close interconnection between solar activity indices  is very
important for prognoses of cyclic variations indices studied here
and other important indices (for example 30,4 nm fluxes, that is
very significant for determination of the Earth high thermosphere
levels heating) [2,4].

We have found out the cyclic behavior of yearly values of
correlation coefficients $K_{corr}$ of linear regression between W
and $F_{10,7}$ during 21,22 and 23 solar activity cycles (see Figure
4). We see that yearly values of $K_{corr}$ have the maximum values
at growth and decrease cycle's phases so the linear connection
between indices is more strong in these cases than at minimum and
maximum cycle's phases. It means that the prognoses of solar
indices, based on  $F_{10,7}$  observations, will be more successful
at growth and decrease cycle's phases.

The yearly values of $K_{corr}$ cyclic variations  are characterized
by cyclic variations with the period T that is equal to half path of
11-year period (5,5 years about). Our study of linear regression
between W and $F_{10,7}$ confirms the fact that at minimum and at
maximum cycle's phases the nonlinear state of interconnection
between solar activity indices (characterized the different levels
of solar atmosphere) increases.

Acknowledgements. The authors thank the RFBR grant 09-02-01010 for
support of the work.

\vskip12pt {\bf REFERENCES}
\vskip12pt

1. L. Svalgaard, H. S. Hudson, Publication: SOHO-23: Understanding a
Peculiar Solar Minimum ASP Conference Series Vol. 428, proceedings
of a workshop held 21-25 September 2009 in Northeast Harbor, Maine,
USA. Edited by Steven R. Cranmer, J. Todd Hoeksema, and John L.
Kohl. San Francisco: Astronomical Society of the Pacific, P. 325
(2010).

2. J. Skupin, M. Weber, H. Bovensmann, and J.P. Burrows,  Proc. of
the 2004 Envisat and ERS Symposium (ESA SP-572), 6-10 September
2004, Salzburg, Austria. Edited by H. Lacoste (2005).

3. R.A. Viereck, L.C. Puga, D. McMullin, D. Judge, M. Weber, and W.
Tobiska, Geophys. Res. Lett,{\bf 28}, P. 1343 (2001).

4.  E.A. Bruevich, A.A. Nusinov, Geomagnetizm i Aeronomiia, {\bf
 24}, P. 581, In Russian, (1984).

5. V.N. Ishkov, Proc. symposium working meeting The activity cycles
on the Sun and stars, Moscow, 18-10 December, Edited by EAAO,
St-Petersburg, P. 57, In Russian, (2009).

6. R.U. Lukyanov, K. Mursula, Proc. symposium working meeting The
activity cycles on the Sun and stars, Moscow, 18-10 December, Edited
by EAAO, St-Petersburg, P. 153, In Russian, (2009).

7. E.A. Bruevich, E.V. Kononovich, Vestn. Mosk. Univ. Fiz. Astron.,
No. 1, P. 70 (2011) [Moscow University Physics Bulletin, {\bf 66},
No. 1, P. 72 (2011)], arXiv:1102.3976v1 (2011).

8. S.L.Baliunas, R.A.Donahue, W.H.Soon, Astrophys.J. {\bf 438},
 P. 269 (1995).

9.E.A. Bruevich, M.M. Katsova, and D.D. Sokolov, Astron. zh. {\bf
78}, P. 827 (2001).

10. E.A. Bruevich, I.Yu. Alekseev, Astrophysics, {\bf 50}, No. 2, P.
 187 (2007), arXiv:1012.5527v1 (2010).

\bigskip

\end{document}